\documentclass[11pt]{article}

\usepackage{epsfig,graphicx}
\usepackage{latexsym}

\parskip =\medskipamount
\thicklines
\def\be{\begin{equation}}
\def\bea{\begin{eqnarray}}
\def\ee{\end{equation}}
\def\eea{\end{eqnarray}}

\newcounter{fig}

\def\d{\mbox{d}}

\def\fp{\displaystyle}
\def\ba{\begin{array}{rcl}}
\def\ea{\end{array}}

\def\o0{\overrightarrow{0}}

\begin{document}

\begin{center}
{\Large \bf Arithmetic area for $m$ planar Brownian paths }\\[0.5cm]

{\large \bf Jean Desbois}\footnote{jean.desbois@u-psud.fr} and 
{\large \bf St\'ephane Ouvry}\footnote{stephane.ouvry@u-psud.fr}\\[0.1cm]
Universit\'e Paris-Sud, Laboratoire de Physique Th\'eorique et Mod\`eles
Statistiques\footnote{Unit\'e Mixte de Recherche CNRS-Paris Sud, UMR 8626}\\
91405 Orsay, France
\\[0.2cm]
\end{center}

\begin{abstract}
We pursue the analysis made in \cite{DesOu} on the arithmetic area enclosed by $m$ closed Brownian paths. We pay a particular attention to the random variable 
$S_{n_1,n_2,\ldots,n_m}(m)$ which is the arithmetic area of the set of points, also called winding sectors, enclosed $n_1$ times by path $1$, $n_2$ times by path $2,\ldots,n_m$ times by path $m$. Various results are obtained in the asymptotic limit $m\to\infty$. A key observation  is that, since the  paths are independent, one can use in the $m$ paths case the SLE information, valid in the $1$-path case,   on the $0$-winding sectors arithmetic area.
\end{abstract}

\section{Introduction}
In \cite{DesOu}, the asymptotic behavior  of the average arithmetic area    enclosed by the external frontier of  $m$ independent closed  Brownian planar paths,  of  same  length $t$ and starting from and ending at the same  point, has been obtained using a path integral approach \cite{Feyn, Sn}. 

In the one path case,
the random variable of interest happens to be  the arithmetic area $S_n$ of the $n$-winding sectors enclosed by  the path,  from which the total arithmetic area  $S=\sum_n S_n$  can be computed.
A $n$-winding sector is by definition a set of points enclosed $n$
 times by the path.    Path integral technics \cite{Sn} give $\langle S_n\rangle= t/(2\pi n^2)$ but end up being  somehow less adapted for $\langle S_0\rangle$, the average arithmetic area of the  $0$-winding sectors inside the path, i.e. of the set of points enclosed an equal number of times clockwise and anti-clockwise by the path. Indeed  path integral  cannot distinguish   $0$-winding sectors inside  the path from the  outside of the path, which is also $0$-winding.  Other technics have to be used, in the case at hand SLE technics
 \cite{Jeunes}, to get $\langle S\rangle=\pi t/5$ from which $\langle S_0\rangle=\pi t/30$ can be  derived. It means that $\langle S_0\rangle=q \langle S-S_0\rangle$  with $q=1/5$. 
 
\begin{figure}
\begin{center}
\includegraphics[scale=.40,angle=0]{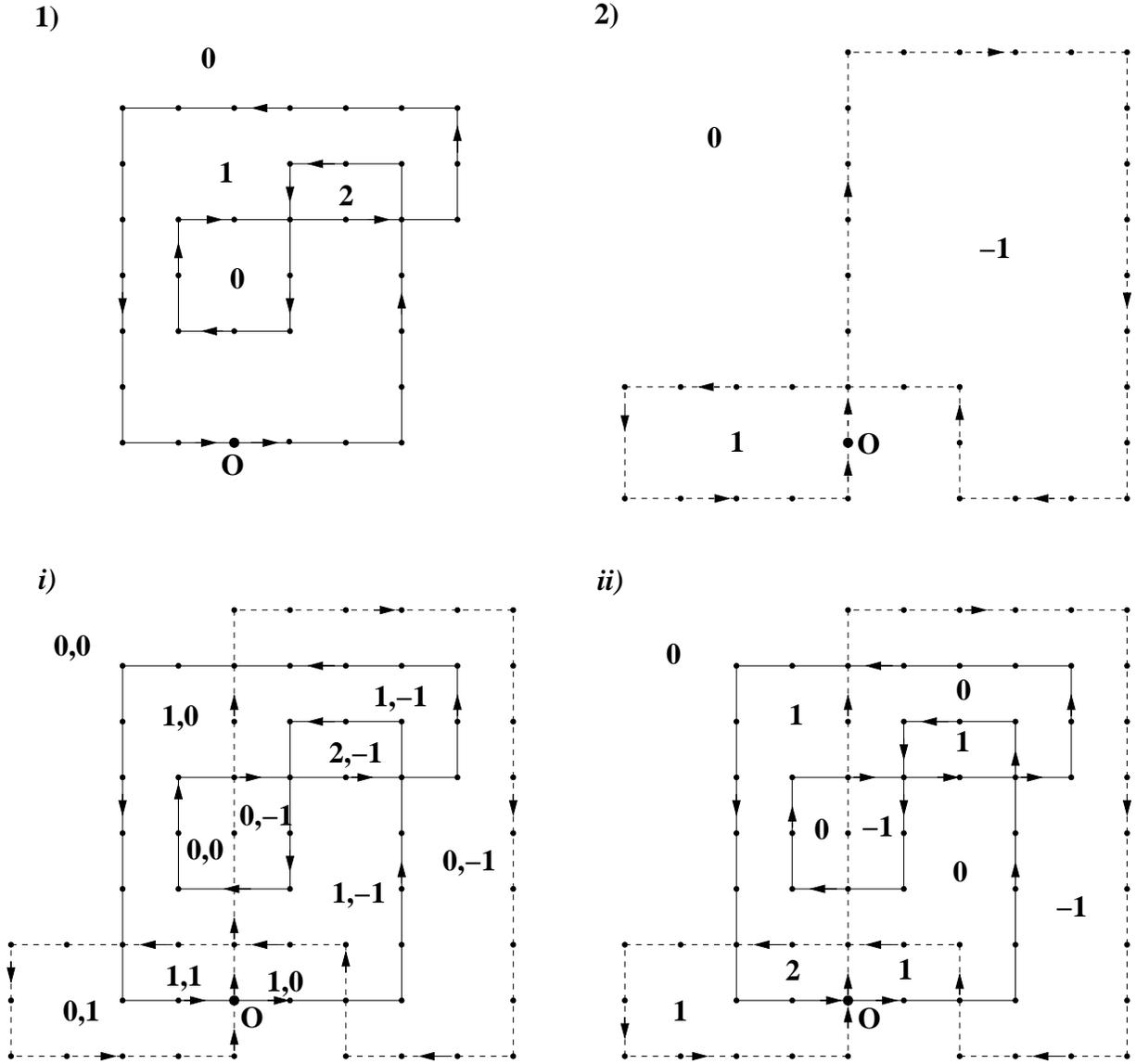}
\caption{On a square lattice two independent closed walks $1$ and $2$  with $38$ steps each, starting from and returning at the origin.  The winding sectors of each walk are labelled by their winding numbers $\{n_1\}$ (for  walk $1$)  and $\{n_2\}$ (for walk $2$). The winding sectors enclosed by the external frontier of the two walks $1 + 2$ are labelled  i) by their joint winding numbers $\{n_1, n_2\}$, ii)   by their total winding number $\{n_1+ n_2\equiv n \}$. The  $n=0$-winding sectors of interest, namely inside the external frontier of $1+2$, correspond  either to $\{0, 0\}$ with at least one of the $0$-winding sectors inside one of the paths, or to $\{n_1, n_2\}$ with $n_1+n_2=0$ and both $n_1$ and $n_2\ne 0$.}
\label{f0}
\end{center}
\end{figure}

  For $m$ independant paths,  the same path integral technics used in the one path case have now to focus \cite{DesOu} on  the random variable $S_n(m)$, the  arithmetic area of the $n$-winding sectors enclosed by the $m$ paths, from which the total area $S(m)=\sum_n S_n(m)$ can in principle be computed.  A  $n$-winding sector is again defined as a set of points enclosed $n$
 times by the $m$ paths 
 as illustrated in Figure \ref{f0} in the  $m=2$ paths case.  One has found that the leading asymptotic term scales like $\ln m$, namely that $\langle S(m)-S_0(m)\rangle\sim {\pi t\over 2}\ln m$ with, as already stressed,  no information on $\langle S_0(m)\rangle$  inside the $m$ paths.

In \cite{Julien} on the other hand,  the asymptotic behavior of the arithmetic area enclosed by the convex envelop of $m$ closed paths was  found to be  $\langle S(m)\rangle_{\rm convex}\sim  {\pi t\over 2}\ln m$, i.e. the same scaling as  $\langle S(m)-S_0(m)\rangle$.  One then concluded \cite{DesOu} that, necessarily, $\langle S(m)\rangle\sim  {\pi t\over 2}\ln m $  and that  $\langle S_0(m)\rangle$ is  subleading.

This particular scaling might be of interest for spatial ecology considerations where one  asks  about the increase in size of a natural reserve with the animal population (assuming of course that  food is homogeneously  and sufficiently available). By identifying an animal looking for food to a random walker one thereby obtains the $\log m$ scaling rather than the naive geometrical $\sqrt{m}$ scaling.

In the present work we will give a more detailed analysis  of the random variable $S_n(m)$. In particular we will  pay  attention to the
random variable $S_{n_1,n_2,\ldots,n_m}(m)$, the arithmetic area of the set of points enclosed $n_1$ times by path $1$, $n_2$ times by path $2,\ldots,n_m$ times by path $m$,  with $\sum_in_i=n$.  Happily enough,  this variable can  be tackled by path integral technics analogous to the one used in \cite{DesOu,Sn} provided that at least one of the $n_i\ne 0$. Note, when $\sum_i n_i=0$, that    $S_{n_1,n_2,\ldots,n_m}(m)$ contains a part of the inside $0$-winding sector area $S_0(m)$. To get a  hand on the other part $S_{0,0,\ldots,0}(m)$, with at least one of the $n_i=0$-winding sectors  inside the corresponding path $i$ (otherwise one would be trivially outside the external frontier of the $m$ paths),   one key observation  is that since the paths are independent one can use  in the $m$ paths case the SLE information $\langle S_0\rangle=q \langle S-S_0\rangle$ valid in the one path case.

From these considerations  will follow 
$\langle S(m)\rangle$  and $\langle S_0(m)\rangle$. We will show in particular that, when $m\to\infty$, the subleading $\langle S_0(m)\rangle$ remains finite. Also some information on  the asymptotics of  
 $\langle S_{n_1,n_2,0,\ldots,0}(m)\rangle$  when both $n_1$ and $n_2\to\infty$ will be obtained. Finally $\langle S_0(2)\rangle$ and the  overlap between   $2$  paths, $\langle 2S(1) -S(2)\rangle$, will be considered. This quantity  might have some interest in polymer physics where polymers are modelised by Brownian paths. 
 
In view of unifying  notations between  \cite{DesOu} and \cite{Sn},  we always denote the $n$-winding sectors arithmetic area and total arithmetic area for $m$ paths by $ S_n(m)$ and  $S(m)$, which means that, from now on,  $ S_n(1)$ and  $S(1)$  stand for $ S_n$ and  $S$, the $n$-winding sectors arithmetic area and total arithmetic area in the  one path case.

 \section{Winding sectors}\label{arw}
 
\subsection{Winding angle and propagator}\label{arw1}

  As stated in the Introduction,  the  arithmetic area of the winding sectors   enclosed by  planar Brownian paths
   can be obtained from their winding properties. Consider a path of length $t$,
 starting from and ending    
  at  $\overrightarrow{r}$ and  the angle $\theta $  wound by the
 path around the origin $O$.  The average of the random variable $e^{i \alpha \theta  } 
  $ over the set of such paths  is
\be
 \langle e^{i \alpha \theta  } \rangle  = \frac{{\cal G}_\alpha (\vec r , \vec r)}
   {{\cal G}_0 (\vec r , \vec r)} \qquad  \label{b1}  \ee
    where 
    \be  {\cal G}_\alpha (\vec r , \vec r)  = 
 \int_{\vec r (0) = \vec r}^{\vec r (t) = \vec r}
 {\cal D}\vec r (\tau)  e^{-\frac{1}{2}\int_0^t 
   \dot{\vec r} ^2 (\tau) \d \tau  +i \alpha  \int_0^t  \dot\theta (\tau) \d \tau  } 
  \qquad   \label{bb1}\ee
  is  the quantum propagator of a charged particle coupled to a vortex
 at location $O$. 
 By symmetry,  it depends only on $r$
  \be 
      {\cal G}_\alpha (\vec r , \vec r)   = \frac{1}{2\pi t} e^{-\frac{r^2}{t}} \sum_{k=-\infty}^{+\infty}
  I_{\vert k- \alpha \vert } \left(  \frac{r^2}{t}  \right)
 \qquad    \label{b2} \ee
 where  $I_{\vert k- \alpha \vert }$ is a modified Bessel function
 and
\be        {\cal G}_0 (\vec r , \vec r)      =  \frac{1}{2\pi t}  \qquad   \label{bb2} \ee 
Additional symmetry and periodicity considerations lead to
  ${\cal G}_\alpha  = {\cal G}_{\alpha +1}  = {\cal G}_{1- \alpha } $ so that   $\alpha$
can be restricted   to  $0 \le  \alpha \le 1$. We also  set \  $\fp \frac{r^2}{t}\equiv x $
   so that  
   $$\langle e^{i \alpha \theta  } \rangle = e^{-x} \sum_{k=-\infty}^{+\infty}
  I_{\vert k- \alpha \vert }(  x )\equiv G_\alpha (x) $$ 
  with $G_0(x)=1$. Clearly areas  are proportional to $t$, so 
 we can, without loss of generality, set $t=1$. 

Let us rewrite  $ G_\alpha(x) $ in a more suitable form : one has

\be
   G_\alpha  (x)    =   e^{-x}   \sum_{k=0}^{+\infty}  I_{ k+ \alpha  } (x)+\{\alpha\to 1-\alpha\}  \label{b3} 
   \ee

Observing that  $\fp  \frac{\d  }{\d x}\left (e^{-x}   \sum_{k=0}^{+\infty}  I_{ k+ \alpha  } (x)\right)=\frac{1}{2} e^{-x} 
   \left(   I_{ \alpha  -1  } (x) -  I_{ \alpha } (x)  \right)        $, we get
\be
  \frac{\d  G_\alpha (x)}{\d x} = \frac{1}{2} e^{-x} 
\left(   I_{-\alpha } (x) -  I_{\alpha } (x) +
  I_{ \alpha  -1  } (x) -  I_{1-\alpha } (x)       \right)
 = \frac{\sin (\alpha\pi )}{\pi } \; e^{-x} \left( K_{ \alpha } (x) + K_{ 1-\alpha } (x)
   \right)   \label{dga1}   
\ee
Using the integral representation \cite{abra} of $K_{ \alpha } (x)$
\be\label{ka}
K_{ \alpha } (x)=\frac{1}{2} \int_{-\infty}^{\infty} \d u \;  e^{-x \cosh u} \; \cosh (\alpha u)
\ee
we deduce
\be
  G_\alpha (x)   =  \frac{\sin (\alpha\pi )}{\pi } \;
  \int_{-\infty}^{\infty} \d u \frac{ 1- e^{-x(1+ \cosh u)} }{1+ \cosh u} \; \cosh \frac{u}{2} 
\cosh ((\alpha -\frac{1}{2})u)
 \label{ga} \ee
Clearly 
\be
  \lim_{x\to \infty}G_\alpha (x)    =\frac{\sin (\alpha\pi )}{\pi } \;    
 \int_{-\infty}^{\infty} \d u \frac{ 1 }{1+ \cosh u} \; \cosh \frac{u}{2} 
\cosh ((\alpha -\frac{1}{2})u)=1 \qquad  \label{ga1} \ee
so that
\be
1-G_\alpha (x)  = \frac{\sin (\alpha\pi )}{\pi } \;
  \int_{-\infty}^{\infty} \d u \frac{ e^{-x(1+ \cosh u)} }{1+ \cosh u} \; \cosh \frac{u}{2} 
\cosh ((\alpha -\frac{1}{2})u)
 \qquad  \label{ga2} 
\ee
Eq. (\ref{ga2}) will be extensively used in the following.

\subsection{Average area  $\langle  S_n(m) \rangle $ of $n$-winding sectors}\label{snm}

Let us first  consider the average arithmetic area $\langle  S_n(m) \rangle $ of 
 the $n$-winding sectors  labelled by their winding number $n$.  
 For $m$ brownian paths of same length
 unity, starting from and ending at the same point, one has \cite{DesOu}
\be\label{zam} 
Z_\alpha(m)\equiv \pi \int_0^\infty \d x \left(1-(G_\alpha (x))^m \right)=
  \sum_{n \ne 0} \langle  S_n(m) \rangle  \left(1- e^{i  2 \pi \alpha  n} \right)
\ee
Rewriting $G_{\alpha}(x) = 1-(1-G_{\alpha}(x)) $  leads,  for $n \ne 0$, to
\be\label{zam1}
\langle  S_n(m) \rangle =-\int_0^1 \; \d \alpha \; Z_\alpha(m)  \cos(2 \pi \alpha n)
=\sum_{j=1}^m (-1)^{j+1}  { m \choose j }  \left(-\pi \int_0^1 \d \alpha \int_0^\infty \d x \left(1-G_\alpha (x)\right)^j \cos(2 \pi \alpha n)\right)
\ee
where $ { m \choose j } $ is the binomial coefficient. Thus
\be\langle  S_n(m) \rangle 
=\sum_{j=1}^m (-1)^{j+1}  { m \choose j }I_{n,j}\ee
with
\be I_{n,j}=-\pi \int_0^1 \d \alpha \int_0^\infty \d x \left(1-G_\alpha (x)\right)^j \cos(2 \pi \alpha n)\ee
Using (\ref{ga2}) one rewrites $I_{n,j}$ as 
\be
  I_{n,j}
      = - \pi\left(\prod_{i=1}^j\int_{-\infty}^{\infty} {\frac{\d u_i}{2\pi  \cosh \frac{u_i}{2}}} \right)
  \frac{\Phi_{n,j}}{j+\sum_{i=1}^j \cosh u_i   }   
\qquad   \label{inj1}
\ee
 where the $ \Phi_{n,j} $'s follow
 from the  integration  over $\alpha$
\begin{itemize}
\item {$j$ odd $= 2k+1$} 
\be \Phi_{n,j}   = 
  \frac{(-1)^{k+1}}{2^{2k+1}} 2\pi \cosh \left(\frac{\sum_{i=1}^j u_i}{2} \right)\sum_{N=n-k}^{n+k+1} (-1)^{N+k-n} 
{2k+1 \choose N+k-n} \frac{2N-1}{ (\pi (2N-1))^2 + (\sum_{i=1}^j u_i)^2}
    \label{phi1}    \ee
 \item   {$j$ even $= 2k$} 
\be \Phi_{n,j}     =  
  \frac{(-1)^k}{2^{2k}} 2(\sum_{i=1}^j u_i )\sinh \left(\frac{\sum_{i=1}^j u_i}{2}\right) \sum_{N=n-k}^{n+k} (-1)^{N+k-n} 
{2k \choose N+k-n} \frac{1}{ ( 2 \pi N)^2 + (\sum_{i=1}^j u_i)^2}
   \label{phi2}
\ee
\end{itemize}
It is possible to   compute $ I_{n,1}$  exactly \cite{Jeunes}   
\be\label{inn1}
I_{n,1}=\frac{\pi}{2} \int_{-\infty}^{\infty} \frac{\d u}{1+\cosh u}\left(    
 \frac{2N-1}{ (\pi (2N-1))^2 + u^2} - \frac{2N+1}{ (\pi (2N+1))^2 + u^2} \right)=
 \frac{1}{2 \pi n^2}
\ee
On the other hand when $n\to\infty$
\begin{itemize}
\item{j  odd  }
 \be  I_{n,j} \sim
  \frac{ (-1)^{\frac{j-1}{2}} }{(2\pi )^j}
  \frac{j!}{n^{j+1}} \; c_j  \label{inas}   \ee
  \be
 c_j =  \int_0^\infty \d x \; e^{-j x} (K_0(x))^j   \label{cj}   \ee
\item{ j even } 
\be I_{n,j} \sim
 \frac{ (-1)^ {(\frac{j}{2}+1)} }{(2\pi )^{j+1}}
  \frac{j \; (j+1)!}{n^{j+2}} \; d_j \label{inas1}   \ee
\be
 d_j =  \int_0^\infty \d x \; e^{-j x} (K_0(x))^{j-1} \int_0^\infty \d u \; e^{- x \cosh u} \; u \;\tanh \frac{u}{2}  \label{cj1} 
\ee
\end{itemize}
We finally obtain
\bea
 \langle  S_n(1) \rangle  &=& I_{n,1}= \frac{1}{2 \pi n^2}  \nonumber \\
 \langle  S_n(2) \rangle    &=& \frac{2}{2 \pi n^2} -I_{n,2}=_{n\to\infty} \;   
  \frac{2}{2 \pi n^2} - \frac{3}{2 \pi^3 n^4} \; d_2 + O(\frac{1}{n^6})    \nonumber   
 \eea
\noindent and in the general case 
  \bea
  \langle  S_n(m) \rangle     &=& \frac{m}{2 \pi n^2}-{m \choose 2} I_{n,2} + 
{m \choose 3} I_{n,3} - \ldots      \nonumber  \\
   &=&_{n\to\infty} \;  \frac{m}{2 \pi n^2} 
  - \frac{3}{4 \pi^3 n^4} \; \left(2 {m \choose 2} d_2+ {m \choose 3} c_3    \right) +
 O(\frac{1}{n^6})  \nonumber
\eea
\noindent with the convention that ${m \choose j}=0$ if $j>m$, $d_2\simeq 2.84 $ and $c_3\simeq 5.73$.
Clearly, and as discussed in the Introduction, we have no information so far on the  $0$-winding sector inside the $m$ paths. To make some progress on this issue, we have  to turn to $\langle  S_{n_1,n_2,...,n_m}(m) \rangle $, the average arithmetic area  of the sectors enclosed  $n_1$ times by path $1$, $n_2$ times by path $2, \ldots,$ and $n_m$ times by path $m$.  
\subsection{Average area $\langle  S_{n_1,n_2,...,n_m}(m) \rangle $}\label{sn1n2}

Winding sectors can as well be  labelled by the set $\{n_1,n_2,\ldots,n_m   \}$ of the individual winding numbers $n_i$ enclosed by each  path $i$. 
 In line with  section (\ref{snm}), one has

\be\label{za1a2}
Z_{\alpha_1,\alpha_2,...,\alpha_m}(m)\equiv \pi \int_0^\infty \d x 
   \left(1- \prod_{i=1}^m G_{\alpha_i}(x)\right)= \sum\; ' \langle 
 S_{n_1,n_2,...,n_m}(m) \rangle
 \left(1- e^{2 i\pi (\sum_{i=1}^m \alpha_i n_i  )} \right)
\ee
where in $\sum'$ the set $n_1=n_2= ...=n_m=0$  is excluded from  the sum.
So
\bea
 \langle  S_{n_1,n_2,...,n_m}(m) \rangle &=& - \int_0^1 \d \alpha_1 ...\int_0^1 \d \alpha_m \;
 Z_{\alpha_1,\alpha_2,...,\alpha_m}(m) \; \cos \left( 2 \pi 
   \left(\sum_{i=1}^m \alpha_i n_i  \right)    \right) 
 \nonumber \\
 &=&  - \int_0^1 \d \alpha_1 ...\int_0^1 \d \alpha_m \; 
 Z_{\alpha_1,\alpha_2,...,\alpha_m}(m) \; \prod_{i=1}^m   \cos(2 \pi \alpha_i n_i )\quad\quad \label{toto}
\eea
where we have used 
  $\int_0^1 \d \alpha \sin (\pi \alpha ) \sin (2 \pi \alpha n ) \cosh \left( (\alpha -\frac{1}{2}) u\right)
 =0  $.
Now $ \langle  S_{n_1,n_2,...,n_m}(m) \rangle $ is invariant by permutation on the $n_i$'s, so one can focus without loss of generality on
$ \langle  S_{n_1,...,n_j,0,...,0 }(m) \rangle $, $1 \le j \le m$, with $n_1,...,n_j\ne 0$.

 Rewriting   $G_{\alpha_i}(x) = 1-(1-G_{\alpha_i}(x)) $  in (\ref{za1a2},\ref{toto}) leads to consider when $n_i \ne 0$
\be
 \int_0^1 \d \alpha_i \; (1-G_{\alpha_i} (x))  \cos (2 \pi \alpha_i n_i ) = \int_{-\infty}^{\infty} \d u_i \; 
 e^{-x(1+\cosh u_i)}  P(u_i,n_i)    \nonumber  \ee
 with

\be P(u_i,n_i)      =  \frac{u_i^2+(1-4 n_i^2)\pi^2}{ (u_i^2+(\pi (2n_i+1))^2)  (u_i^2+(\pi (2n_i-1))^2) }\sim_{n_i\to\infty}-\frac{1}{4 \pi^2 n_i^2}
 \ee
It follows that
\be\label{sn1n200}
 \langle  S_{n_1,...,n_j,0,...,0 }(m) \rangle = \pi (-1)^j \; \int_0^\infty \d x 
\left( \prod_{i=1}^j \; \int_{-\infty}^{\infty} \d u_i \; e^{-x(1+\cosh u_i)}\; P(u_i,n_i) \right) (1-f(x))^{m-j}
\ee
where
\be\label{fx}
 \qquad f(x)= \int_0^1  \d \alpha \;
  (1-G_\alpha (x)) = \int_{-\infty}^{\infty} \d u \; \frac{e^{-x(1+\cosh u)}}{u^2+\pi^2}
\ee  
 For example when $m=1, 2$ one gets
\bea
  \langle  S_{n_1}(1) \rangle    &=&
  - \pi \int_0^\infty \d x  \int_{-\infty}^{\infty} \d u \; 
       e^{-x(1+\cosh u)} P(u,n_1)=\frac{1}{2\pi n_1^2} \nonumber   \\
 \langle  S_{n_1,n_2}(2) \rangle    &=&  
\pi \int_{-\infty}^{\infty}\int_{-\infty}^{\infty}  \frac{ \d u_1 \;  \d u_2 }{2+\cosh u_1 +\cosh u_2}  P(u_1,n_1) P(u_2,n_2)
       \nonumber   \\
    \qquad  \langle  S_{n_1,0}(2) \rangle    &=& \frac{1}{2\pi n_1^2}+\pi   
 \int_{-\infty}^{\infty}\int_{-\infty}^{\infty}  \frac{ \d u_1 \;  \d u_2 }{2+\cosh u_1 +\cosh u_2}  P(u_1,n_1) P(u_2,0) 
    \nonumber  
\eea
When  $n_i\to\infty$ one obtains
\bea
 \langle  S_{n_1,...,n_j,0,...,0 }(m) \rangle     &\sim &   \frac{1}{2^j \pi^{2j-1}}   
   \frac{1}{n_1^2 ... n_j^2}
c_{j,m} \nonumber       \eea
where the constant 
\bea
  c_{j,m}   &= & \int_0^\infty \d x \; e^{-j x} \; (K_0(x))^j \;(1 -f(x))^{m-j}
               \nonumber      
\eea
has to be evaluated  numerically to the exception of  


\[c_{2,2}={3\over 2} \zeta(2)-1+{3\over 2} \zeta(2) \sum_{k=1}^{\infty}\prod_{i=1}^k(1 - {1\over 2 i})^2 - 
  \sum_{k=1}^{\infty}\prod_{i=1}^k(1 - {1\over 2 i + 1})^2 \]

\section{Arithmetic area enclosed by $m$ Brownian paths}

 We are now in position to compute     $\langle  S(m) \rangle $, the average arithmetic area
 enclosed by the $m$  paths. Obviously
\be\label{sm}
\langle  S(m) \rangle = \sum  \; ' \langle  S_{n_1,...,n_m}(m) \rangle + 
  \langle  S_{0,...,0}(m) \rangle
\ee
where $ \langle  S_{0,...,0}(m) \rangle$ is the area of the  finite $\{n_1=0,n_2=0,\ldots,n_m=0\}$ winding sectors  inside, at least, one of
 the paths\footnote{Remember that
$\langle  S_0(m) \rangle =\langle  S_{0,...,0}(m) \rangle+\sum' \langle  S_{n_1,...,n_m}(m) \rangle $, with $\sum n_i=0$  and at least one of the $n_i\ne 0$.}.
From (\ref{za1a2}), we get
\be\label{sm1}
  \sum \; ' \langle  S_{n_1,...,n_m}(m) \rangle=\int_0^1 \d \alpha_1 ...\int_0^1 \d \alpha_m
 \; Z_{\alpha_1,...,\alpha_m}(m)
\ee
so that, using  (\ref{fx}),
\be\label{sm2}  
 \sum \; ' \langle  S_{n_1,...,n_m}(m) \rangle \; = \; \pi \; \int_0^\infty \; \d x \;
 \left( 1- (1-f(x))^m \right)
\ee

In order to find   $ \langle  S_{0,...,0}(m) \rangle $ first consider
$$
 \sum_{n_1 \ne 0} \langle  S_{n_1,0,...,0}(m) \rangle \;  = \; \pi \; \int_0^\infty  \; \d x \; f(x)
 \left( 1-f(x) \right)^{m-1}
$$
and, the sum
\bea
A_1 & \equiv &  \sum_{n_1 \ne 0} \left( \langle  S_{n_1,0,...,0}(m) \rangle +
  \langle  S_{0,n_1,0...,0}(m) \rangle + ... +\langle  S_{0,...,0,n_1}(m) \rangle     \right)\nonumber \\
 &  = &    {m \choose 1   } \pi \; \int_0^\infty  \; \d x \; f(x)
\left(1-f(x)\right)^{m-1}\nonumber 
\eea
where one has  taken into account  permutation invariance.
Similarly one can consider
\bea
A_2 & \equiv &  \sum_{n_1,n_2 \ne 0} \left( \langle  S_{n_1,n_2,0,...,0}(m) \rangle + ...+ 
  \langle  S_{0...,0,n_1,0...,0,n_2,0...0}(m) \rangle + ...+
\langle  S_{0,...,0,n_1,n_2}(m) \rangle     \right)    \nonumber \\
   &  =  &    {m \choose 2   } \pi \; \int_0^\infty  \; \d x \; f(x)^2
 (1-f(x))^{m-2}    \nonumber \eea
{\rm and in general}\\
\bea A_m  \equiv   \sum_{n_1,n_2...,n_m \ne 0} \;  \langle  S_{n_1,n_2,...,n_m}(m) \rangle
 \;& =& \; {m \choose m   } \pi \; \int_0^\infty  \; \d x \; f(x)^m
   (1-f(x))^{m-m}    \nonumber
\eea
with, obviously, 
 \be\fp \sum_{i=1}^m A_i =  \pi \; \int_0^\infty  \; \d x \;  \left( 1- (1- f(x))^m \right)=
  \sum \; '   \langle  S_{n_1,n_2,...,n_m}(m) \rangle \label{ach}   \ee
We are interested in $ \langle  S_{0,...,0}(m) \rangle $ : in the case of one closed  path one knows from SLE \cite{Jeunes} that
 \be\label{q}
 q=\frac{\langle  S_0(1) \rangle }{ \sum_{n \ne 0} \langle  S_n(1) \rangle  }
 =\frac{1}{5}
\ee
This means that for any point {\bf inside } the path, with winding  number $n$, $q$ is the ratio of the probability to have a $0$-winding to the probability to have a $n\ne 0$-winding.
 
  $A_1$  counts 
 the points with only one non-zero winding number. It follows that the corresponding contribution to  $ \langle  S_{0,...,0}(m) \rangle $  is necessarily
 \ $ q \; A_1$. Similarly, 
 $A_2$ counts the points with only two non-zero winding numbers. Since the $m$ paths are 
 independent, it follows that the corresponding contribution to
  $ \langle  S_{0,...,0}(m) \rangle $ is
 \ $ q^2 \; A_2$.  This line of reasoning generalizes to  $A_k$ : the contribution to $ \langle  S_{0,...,0}(m) \rangle $ is $ q^k \; A_k$.
  Finally  
\be
 \langle  S_{0,...,0}(m) \rangle =\sum_{i=1}^m q^i A_i = \Phi_0(m) - \Phi_q(m)
   \nonumber  \ee
   with
   \be\label{phi}
 \Phi_q(m) = \pi \int_0^\infty \d x 
 \bigg( 1- (1-(1-q)f(x))^m \bigg)
\ee

Clearly $\sum_{i=1}^m A_i$ in (\ref{ach}) coincides with $\Phi_0(m)$. It follows that the average arithmetic area enclosed by the $m$ paths is 
\be\label{aver}
\langle  S(m) \rangle  =  \sum_{i=1}^m (1+q^i) A_i =2 \Phi_0(m) - \Phi_q(m)
\ee
When $m\to \infty$ (see eq.(\ref{phiq}) in the  appendix) one obtains
\be\label{aver1}
\langle  S(m) \rangle  =  \frac{\pi}{2} \ln m - \frac{\pi}{4} \ln \ln m
 + \frac{\pi}{2} \left( \ln \sqrt{\frac{25}{4 \pi^3}}  + C \right) +o(\frac{1}{\sqrt{\ln m }})
\ee  
where $C$ is the Euler constant.
\begin{figure}
\begin{center}
\includegraphics[scale=.40,angle=0]{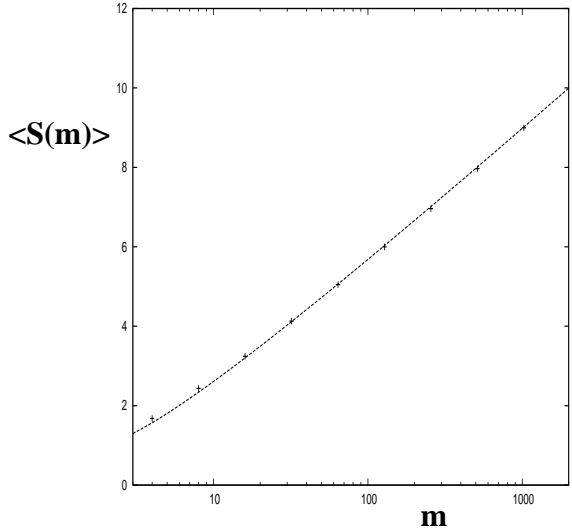}
\caption{The average arithmetic area $\fp   \langle S(m)  \rangle $: the crosses are 
numerical simulations (10000 events) of closed random walks ($10^6$ steps for each one)
 on the 2D square lattice; $m = 4, 8, 16, ..., 1024$; the line is 
    the   analytical result eq.(\ref{aver1}). The agreement is quite correct as soon as 
  $m \ge 16$.
}\label{f2}
\end{center}
\end{figure}
In Figure \ref{f2}, numerical simulations  for $ \langle  S(m) \rangle  $
  show that the agreement with eq.(\ref{aver1}) is quite correct,
 even for not  large values of $m$. 
 
 Moreover, the asymptotic \cite{DesOu}
 of  $ \langle S(m)-  S_0(m)\rangle $  when  $m\to \infty$  is known to be
\be\label{averdo1}
\langle  S(m) -  S_0(m) \rangle  =  \frac{\pi}{2} \ln m - \frac{\pi}{4} \ln \ln m
 + \frac{\pi}{2} \left(- \ln \sqrt{4 \pi }  + C \right) +o(\frac{1}{\sqrt{\ln m }})
\ee 
Comparing Eqs.(\ref{aver1}) and (\ref{averdo1}), one deduces that the subleading $0$-winding sector  average arithmetic area  
\be\label{aver0}
\lim_{m\to\infty}\langle  S_0(m )\rangle = \frac{\pi}{2} \ln \frac{5}{\pi} 
\ee 
remains finite in the $m\to\infty$ limit.
 

Finally, as another illustration of the path integral formalism, consider  the  average  overlap $\langle 2S(1)- S(2) \rangle $ of the arithmetic areas
of two paths  and $\langle S_0(2) \rangle $, the average $0$-winding sectors arithmetic area of two  paths. One has
\bea
    \langle S(1) \rangle &=& \pi \int_0^\infty \d x (1+q) \; f(x)
       =\frac{\pi }{5}       \nonumber \\
  \langle S(2) \rangle &=&  \pi \int_0^\infty \d x 
\left( 2(1+q)f(x)+f(x)^2((1-q)^2-2)  \right)  \nonumber \eea
so that 
\bea\langle 2S(1)- S(2) \rangle      &=& 
  \pi (2-(1-q)^2)      \int_0^\infty \d x  \; f(x)^2       \nonumber \\
       &=&  \frac{34 \pi }{25}  \int_{-\infty}^{\infty}\int_{-\infty}^{\infty} 
\frac{\d u_1 \d u_2}{(2+\cosh u_1+\cosh u_2)(\pi^2+u_1^2)(\pi^2+u_2^2)}    \label{achach} 
\eea
 Numerically 
 $$\fp \frac{\langle 2S(1)- S(2) \rangle }{\langle S(1) \rangle} \approx 0.286$$ 
 is
 close to what one would obtain if the paths   attached in $O$
 were two circles of  radius $R$:  the overlap in unit of $\pi R^2$ would then be 
 $$\frac{1}{2}-\frac{2}{\pi^2} \approx 0.297$$

Also, as far as  $\langle  S_0(2) \rangle$   is concerned,  rewrite
   $$ \langle  S_0(2) \rangle =\langle  S(2) \rangle-  \sum_{n \ne 0}  \langle  S_n(2) \rangle =
  2 \langle S(1)\rangle-  \langle 2S(1)- S(2) \rangle  -  \sum_{n \ne 0}  \langle  S_n(2) \rangle  $$ 
  From Section \ref{snm}  one has 
$$
\sum_{n \ne 0}  \langle  S_n(2) \rangle = \frac{\pi}{3} - \pi   \int_{-\infty}^{\infty}\int_{-\infty}^{\infty}
\d u_1 \d u_2 \frac{\tanh \frac{u_1}{2} + \tanh \frac{u_2}{2}}
{(2+\cosh u_1+\cosh u_2)(u_1+u_2)((2 \pi)^2+(u_1+u_2)^2)}  
$$
so that, using  $\fp  \langle  S_0(1) \rangle =\frac{\pi}{30} $  and eq. (\ref{achach}),  one finds $\langle  2S_0(1)-S_0(2) \rangle$  to be 
\be\label{avs02}
 \pi  \int_{-\infty}^{\infty}\int_{-\infty}^{\infty} 
\frac{\d u_1 \d u_2}{(2+\cosh u_1+\cosh u_2)} 
\left( \frac{\frac{34}{25}}{(\pi^2+u_1^2)(\pi^2+u_2^2)} -
\frac{\tanh \frac{u_1}{2} + \tanh \frac{u_2}{2}}{(u_1+u_2)((2 \pi)^2+(u_1+u_2)^2)} 
\right)
\ee
 If the paths were not overlapping, then one would necessarily have $\langle  2S_0(1)-S_0(2) \rangle=0$. The non vanishing  result (\ref{avs02}) clearly indicates that the two paths do, on average,  overlap as already seen in (\ref{achach}). Note that the $0$-winding sectors "overlap" $\langle 2S_0(1)- S_0(2) \rangle$  is different in nature from the  arithmetic area overlap $\langle 2S(1)- S(2) \rangle$: the latter is a purely  geometric overlap, whereas the former is more subtle since the superposition of the two paths destroy some original $0$-winding sectors in each of the paths and create new $0$-windings sectors for the two paths.

\section{Conclusion}

Path integral technics  have been extensively used to tackle the issue of $n$-winding sectors arithmetic area of $m$ Brownian paths. 
Some information stemming from SLE technics  valid only in the one path case have also proved useful in the $m$ paths case, merely because the paths are independent.

Eqs.(\ref{aver1}) and (\ref{aver0}) are the main results of this paper. In particular the subleading $0$-winding arithmetic area has been shown to remain finite in the asymtotic limit.
Numerical simulations have nicely confirmed these asymptotics results. The overlap between  two paths is also computed numerically. Applications to polymer physics will be studied in a forthcoming publication.

\section{Appendix }
In the appendix we derive the  $m \to \infty $ asymptotic  limit of   $ \Phi_q(m) $ defined in (\ref{phi})
\[
  \Phi_q(m) = \pi \int_0^\infty \d x \left( 1- (1-(1-q)f(x))^m \right) \]
  with $f(x)$  given in (\ref{fx})
\[ f(x) = \int_{-\infty}^{\infty} \d u \frac{e^{-x(1+\cosh u)}}{\pi^2+u^2} 
        \]
Setting $\fp x= \frac{y \ln m}{2}$, $ \Phi_q(m) $ becomes
\bea
   \Phi_q(m) &=&  \frac{\pi  \ln m}{2}  \int_0^\infty \d y 
     \left( 1- (1-(1-q)f( \frac{y \ln m}{2}))^m \right)  \qquad \nonumber  \\
      & \approx & \frac{\pi  \ln m}{2} \int_0^\infty \d y 
  \left( 1-  e^{-m(1-q) f( \frac{y \ln m}{2})   }   \right)  \qquad \label{A04}
\eea
Using  $\fp f(x) \simeq \frac{e^{-2x}}{\pi^2}\sqrt{\frac{2\pi}{x}} $  when $x\to\infty$,   the integrand in (\ref{A04}) 
 $$1- e^{-m(1-q) f( \frac{y \ln m}{2})}  \approx 
 1-e^{-(1-q) \frac{m^{1-y}}{\pi^2}\sqrt{\frac{4\pi}{y \ln m}}} $$
 behaves, in the $m\to\infty$ limit,  like 
\bea  &0&   \qquad \mbox{if}  \qquad y>1   \nonumber\\
                  &1&   \qquad \mbox{if}  \qquad y<1   \nonumber
                  \eea
and so, at leading order,  $\fp  \Phi_q(m) = \frac{\pi  \ln m}{2}$.
                
 Focusing now on the subleading
  correction  at order $\fp \frac{1}{\sqrt{\ln m}}$, namely 
  \be 
 \frac{\pi  \ln m}{2}\left( \int_1^{\infty}\d y \left(1-e^{-(1-q) \frac{m^{1-y}}{\pi^2}\sqrt{\frac{4\pi}{y \ln m}}}\right)  -\int_0^1\d y e^{-(1-q) \frac{m^{1-y}}{\pi^2}\sqrt{\frac{4\pi}{y \ln m}}} \right)  
  \ee
  let us first consider the $y>1$ integration. One has to compute  
$$
a \approx \frac{\pi  \ln m}{2} \int_1^\infty \d y (1-q)\frac{m^{1-y}}{\pi^2}
 \sqrt{\frac{4\pi}{y \ln m}}
 \approx \frac{\pi  \ln m}{2} \int_1^\infty \d y (1-q)\frac{m^{1-y}}{\pi^2}
 \sqrt{\frac{4\pi}{ \ln m}} 
$$
where one has used that, because of the $m^{1-y}$ factor, $y$ is peaked to $1$ when $m\to\infty$.
 One obtains 
\be\label{Aa}
a=\frac{1-q}{\sqrt{\pi \ln m}} +o(\frac{1}{\sqrt{ \ln m}})
\ee

Considering next the $y<1$ integration one has to compute
\bea
b &\approx & -\frac{\pi  \ln m}{2} \int_0^1 \d y
 e^{-(1-q) \frac{m^{1-y}}{\pi^2}\sqrt{\frac{4\pi}{y \ln m}}} \nonumber \\
 &\approx & -\frac{\pi  \ln m}{2} \int_0^1 \d y
 e^{-(1-q) \frac{m^{1-y}}{\pi^2}\sqrt{\frac{4\pi}{ \ln m}}} \nonumber \\
 & =  & -\frac{\pi  \ln m}{2} \int_0^1 \d y e^{-a' m^{1-y}} 
 \qquad \mbox{ with } \qquad  a'=\frac{1-q}{\pi^2} \sqrt{\frac{4 \pi}{\ln m}}       \nonumber
\eea  
Setting $a' m^{1-y}= z$ \  with $\fp -\d y \ln m= \frac{\d z}{z}$ leads to
$$
b \approx -\frac{\pi }{2} \int_{a'}^{ a' m } e^{-z} \frac{\d z}{z}=-\frac{\pi }{2}
 \left(   \left[ \ln z \; e^{-z}\right]_{a'}^{a' m }+ \int_{a'}^{ a' m } \ln z \; e^{-z} \d z
\right)
$$
At order  $\fp \frac{1}{\sqrt{\ln m}}$, one has
 $  \left[ \ln z \; e^{-z}\right]_{a'}^{a' m } \approx -\ln a' \; 
 (1-a') $ and $\int_{a'}^{ a' m } \ln z \; e^{-z} \d z \approx  
\int_0^\infty \ln z \; e^{-z} \d z-\int_0^{a'} \ln z \; e^{-z} \d z  
 \approx C-\int_0^{a'} \ln z \; (1-z) \d z     $  where $C$   is the Euler constant.
  The last remaining integral  
 is straightforward and finally
\be\label{Ab}
 b \approx \frac{\pi }{2} \left( \ln a' +C -a'   \right)
\ee
Noticing that $\fp a'\frac{\pi }{2}=\frac{1-q}{\sqrt{\pi \ln m}}\approx a$, we are left with
 $\fp a+b \approx \frac{\pi }{2} \left( \ln a' +C  \right) $ so that
\be\label{phiq}
 \Phi_q(m)= \frac{\pi }{2} \ln m - \frac{\pi }{4} \ln\ln m +\frac{\pi }{2}
\left(    \ln (1-q) +\ln \sqrt{\frac{4}{\pi^3}} + C
 \right) +o(\frac{1}{\sqrt{\ln m }})
\ee

\end{document}